\documentclass[aps,prd,preprint]{revtex4}
\usepackage{graphicx}
\begin{document}
\draft
\title{ANOMALOUS PRODUCTION OF FOURTH FAMILY UP QUARKS AT FUTURE LEPTON HADRON COLLIDERS}
\author{A.T. Alan$^{1}$, A. Senol$^{1}$ and O. \c{C}ak{\i}r$^{2}$}
\address{$^{1}$ Department of Physics, Abant Izzet Baysal University, 14280, Bolu, Turkey}
\address{$^{2}$ Department of Physics, Ankara University, 06100, Tandogan, Ankara, Turkey}
\begin{abstract}
We investigate the production of fourth family up-type quarks
using effective lagrangian approach at  future lepton-hadron
colliders and study the kinematical characteristics of the signal
with an optimal set of cuts. We obtain the upper mass limits 500
GeV at THERA and one TeV at Linac $\otimes$ LHC.

\end{abstract}
\maketitle
\section{Introduction}
As is well known, the Standard Model (SM) with three families is
in excellent agreement with experimental data available today
\cite{LEP}. But it leaves some open questions. At the most
fundamental level, the number of fermion generations and the
origin of their mass hierarchy are not explained by the SM. For
these reasons, and others, several models extending SM have been
proposed \cite{GR,JDH,MHC,ZE,NE,FP}. Except the minimal SU(5) GUT
all these models accommodate extra fermion generations
\cite{PPM,KTM}.

In the context of the search programs of future colliders, many
analyses have been done for the production of fourth generation
quarks at the linear \cite{VENUS,OPAL,AMY} and at hadron colliders
\cite{BDP}. The potentials of the future lepton-hadron colliders
in the new physics searches are comparable to those of the linear
and hadron colliders \cite{SS}. Thus, in this study, we
investigate the possibility of a single production of a
fourth-family up quark ($u_4$) suggested by the effective
lagrangian approach. In this approach the most general effective
lagrangian, which describes the Flavour Changing Neutral Current
(FCNC) interactions between $u_4$ and ordinary quarks, involving
electroweak boson and gluon is given as follow \cite{H1,TKBX}:
\begin{eqnarray*}
  \mathcal{L}_{eff}&=&\sum_{U=u,c}i\frac{ee_U}{\Lambda}\kappa_{\gamma,u_4}\bar{u}_4\sigma_{\mu\nu}q^{\nu}
  UA^{\mu}+\frac{g}{2\cos\theta_W}\bar{u}_4[\gamma_{\mu}(v_{Z,U}-a_{Z,U}\gamma^{5})
  +i\frac{\kappa_{Z,u_4}}{\Lambda}\sigma_{\mu\nu}q^{\nu}]UZ^{\mu}
\\
  &&+i\frac{g_s}{\Lambda}\kappa_{g,u_4}\bar{u}_4\sigma_{\mu\nu}q^{\nu}\frac{\lambda^i}{2}UG^{i\mu}
  +h.c.,
\end{eqnarray*}
where  $\sigma_{\mu\nu}$=(i/2)[$\gamma^{\mu}$ , $\gamma^{\nu}$ ],
$\theta_W$ is the Weinberg angle, $q$ is the four-momentum of the
exchanged boson; $e$, $g$ and $g_s$ denote the gauge couplings
relative to U(1), SU(2) and SU(3) symmetries, respectively;
$e_{U}$ is the electric charge of up-type quarks, $A^{\mu}$,
$Z^{\mu}$ and $G^{i\mu}$ the fields of the photon, $Z$ boson and
gluon, respectively; and $\Lambda$ denotes the scale up to which
the effective theory is assumed to hold. By convention, we set
$\Lambda$=m$_4$, mass of the fourth family quark in following.

\section{The Anomalous Production of Fourth Family Up Quark}

Parton level subprocess responsible for the $u_4$ production in ep
collisions is $eq\rightarrow eu_4$. The kinematics of this process
is same as that of the single top quark production via FCNC
interactions, which was presented in one of our earlier works
\cite{ATA}. Here, we present the total production cross sections
as functions of $u_4$ mass in Fig.~1 and Fig.~2. In Fig.~1, we
display the cross sections as functions of the mass of $u_4$, at
future lepton-hadron collider THERA with the center of mass energy
$\sqrt{S}$= 1 TeV and with the luminosity of $\mathcal{L}=
4.10^{30}$ cm$^{-2}$s$^{-1}$ \cite{TES}. Fig.~2 shows the
behaviour of the cross section as a function of m$_4$ at Linac
$\otimes$ LHC with $\sqrt{S}$= 5.3 TeV and with the luminosity of
$\mathcal{L}= 10^{33}$ cm$^{-2}$s$^{-1}$ \cite{RB}. In these
figures the lower lines correspond to the photonic channel only
and the upper lines correspond to the sum of $\gamma$ and Z
exchange diagrams. Graphs in Fig~1 and 2 were computed by taking
the anomalous coupling values $\kappa_{\gamma}=\kappa_{Z}=0.1$ for
illustration. In obtaining the mentioned behaviour of the cross
sections we have taken into account of both up and charm quark
distributions \cite{ARWR,JR}. The charm quark is present inside
the proton as part of the quark-antiquark sea and gives
considerable contribution to the cross section.

We assume the anomalous decays of $u_4$ quark to be dominant,
which is different from the case of top-quark decays where the SM
decay mode is dominant. In Table \ref{tab1}, we present the
branching ratios and the total decay widths of $u_4$ quark via
anomalous interactions. SM decay modes are negligible for
$\kappa/\Lambda>0.01$ TeV$^{-1}$ due to the small magnitude of the
extended CKM matrix elements $V_{u_4b}$ \cite{SA,EAOCSS}.

In order to enrich the statistics for the experimental
observations of the signal we also take into account $\bar u_4$
production through the subprocess e$\bar{q}\rightarrow$
e$\bar{u}_4$. The contribution of this process is relatively small
when compared with the $u_4$ production due to the sea quark
distribution in the proton. Tables~\ref{tab2} and~\ref{tab3}
present the total production cross sections of $u_4$ and $\bar
u_4$ in addition to the number of signal and background events in
various decay channels of $u_4$ at THERA and Linac$\otimes$LHC,
respectively.

When the fourth family $u_4$ (or $\bar{u}_4$) quarks are produced,
they will decay via FCNC interactions giving rise to the signal
$e^{-}qV$, where $q=u, c, t$ (or $\bar u,\bar c, \bar t$) and $V$
denotes the neutral gauge bosons $\gamma,Z,g$.

We consider the relevant backgrounds from the following
subprocesses:
\begin{eqnarray*}
&eq\rightarrow eq\gamma &\\
&eq\rightarrow eqZ& \\
&eq\rightarrow eqg&
\end{eqnarray*}
where $q=u,c$ (or $\bar{u},\bar{c}$). The cross sections for these
backgrounds are shown in Table~\ref{tab2} (at THERA) and
Table~\ref{tab3} (at Linac $\otimes$ LHC) for the minimal cuts
$p_T^{e,\gamma,j}>10$ GeV and optimal cuts $p_T^{e,\gamma,j}>20$
GeV and $M_{qV}>250$ GeV on the final state particles. From Table
\ref{tab2} we conclude that the number of signal events for
$u_4\rightarrow gq$ and $u_4\rightarrow Zq$ (q=u,c) channels is
promising, which makes it possible to observe a $u_4$ production
signal at the THERA, especially for low lying $u_4$-quark mass
values (300-500 GeV). As can be seen from Table~\ref{tab3}, it
will be possible to observe the anomalous production of $u_4$
quark in all decay channels if the corresponding backgrounds is
kept at a low level at Linac $\otimes$ LHC collider. We found that
$u_4$ production signal at this machine is observable down to the
anomalous coupling $\kappa_{V,u_4}=0.01$ at the mass of
$u_4$-quark about 700 GeV. For the channels including top quark in
the final state ($e^{-}Vt$) we obtain very low number of
background events, therefore we have not shown them in Tables
\ref{tab2} and \ref{tab3}.

Assuming Poisson statistics, we use the significance formula
$S/\sqrt{B}\geq 3$ for signal observation at the 95\% C.L., where
the number of signal and background events S and B are calculated
by multiplying the cross section with corresponding branching
ratios depending on the decay channels and the integrated
luminosities of the colliders considered.

\section{Conclusion}
In this study, we have considered the anomalous single production
of fourth family up-quarks via the FCNC couplings at future ep
colliders. We have shown that the reaction $eq\rightarrow eu_4$
can take place at an observable rate at these colliders. Hence,
the fourth family up-quark will manifest itself at THERA and Linac
$\otimes$ LHC with masses below 500 GeV and 1 TeV, respectively.
Thus the future lepton-hadron colliders have promising potential
in searching for manifestations of non-standard physics.

\begin{acknowledgements}
This work is partially supported  Abant Izzet Baysal University
Research Found.
\end{acknowledgements}

\newpage
\begin{table}
  \centering
  \caption{Branching ratios ($\%$) and total decay widths of $u_4$ quark depending on its mass ($\kappa_{\gamma,u_4}=\kappa_{Z,u_4}=0.1$,
  $\Lambda=m_{u_4}$)} \label{tab1}

  \begin{tabular}{lccccccc}\hline\hline
    Mass (GeV) & $gu(c)$ & $gt$ & $Zu(c)$& $Zt$ & $\gamma u(c)$ & $\gamma t$ & $\Gamma$(GeV)
    \\\hline
    300 & 3.1 & 0.9 & 70.5 & 25.2 & 0.1 & 0.04 & 7.15 \\
    400& 1.4 & 0.8 & 61.1 & 36.6 & 0.07 & 0.04 & 20.12 \\
    500 & 0.8 & 0.6 & 57.3 & 41.2 & 0.04 & 0.03 & 42.31 \\
    600& 0.5 & 0.4 & 55.2 & 43.8 & 0.03 & 0.02 & 76.29 \\
    700& 0.4 & 0.3 & 53.9 & 45.4 & 0.02 & 0.02 & 124.45
    \\\hline\hline
  \end{tabular}
\end{table}
\vspace{1cm}
\begin{table}
  \centering
  \caption{Number of signal and background events in various
  decay channels of $u_4$ quark at THERA with $\sqrt{S}=1$ TeV and $L=40$ pb$^{-1}$
  with corresponding total cross section in pb.
  B$_1$ and B$_2$ denote the number of background events with
  the cuts ($p_T^{e,\gamma,j}>10$ GeV) and ($p_T^{e,\gamma,j}>20$ GeV, $M_{Vq}>250$ GeV), respectively.}
  \label{tab2}

  \begin{tabular}{lcccccccc}\hline\hline
    Mass (GeV) & $gu(c)$ & $gt$ & $Zu(c)$& $Zt$ & $\gamma u(c)$ & $\gamma t$ & $\sigma_{tot}(eu(c)\rightarrow
    eu_4)$&$\sigma_{tot}(e\bar{u}(\bar{c})\rightarrow
    e\bar{u}_4)$
    \\\hline
    300 & 12.4 & 3.6 & 280.1 & 100.3 & 0.56 & 0.17 &6.67&3.26 \\
    400& 4.1 & 2.2 & 176.6 & 105.8 & 0.20 & 0.10 &4.57&2.66 \\
    500 & 1.7 & 1.2 & 117.8 & 84.6 & 0.08 & 0.06 &3.02&2.12 \\
    600& 0.7 & 0.6 & 75.9 & 60.1 & 0.04 & 0.03 & 1.88&1.55 \\
    700& 0.3 & 0.3 & 43.3 & 36.4 & 0.02 & 0.01 &1.05&0.96 \\\hline
    B$_1$ &$17440$&-&$6.7$&-&$1152$&-&&\\\hline
    B$_2$ &$20.3$&-&$1.3$&-&$41.2$&-&&\\\hline\hline
  \end{tabular}
\end{table}
\vspace{1cm}
\begin{table}
  \centering
  \caption{Number of signal and background events in various decay channels of $u_4$ quark at Linac$\otimes$LHC with $\sqrt{S}=5.3$ TeV and $L=10^4$
  pb$^{-1}$ with corresponding total cross section in pb.  B$_1$ and B$_2$ denote the number of background events with
  the cuts ($p_T^{e,\gamma,j}>10$ GeV) and ($p_T^{e,\gamma,j}>20$ GeV, $M_{Vq}>250$ GeV), respectively.}
  \label{tab3}

  \begin{tabular}{lcccccccc}\hline\hline
    Mass (GeV) & $gu(c)$ & $gt$ & $Zu(c)$& $Zt$ & $\gamma u(c)$ & $\gamma t$ & $\sigma_{tot}(eu(c)\rightarrow
    eu_4)$&$\sigma_{tot}(e\bar{u}(\bar{c})\rightarrow
    e\bar{u}_4)$
    \\\hline
    300 & 6423.1 & 1835.2 & 144692.4 & 51786.5 & 286.7 & 86.0 &13.97&6.54 \\
    400& 2696.1 & 1423.3 & 115137.5 & 68945.8 & 131.1 & 65.5 &13.06&5.78 \\
    500 & 1459.8 & 984.3 & 101194.3 & 72720.7 & 70.9 & 50.0 &12.36&5.29 \\
    600& 890.8 & 681.2 & 91990.9 & 72926.4 & 45.8 & 34.9 & 11.74&4.92 \\
    700& 592.3 & 488.3 & 85088.9 & 71614.7 & 30.4 & 25.4
    &11.15&4.63\\\hline
    B$_1$ &$1.9\times 10^7$&-&$1.2\times 10^{4}$&-&$1.4\times 10^{6}$&-& &\\\hline
    B$_2$ &$6.5\times 10^4$&-&$3.3\times 10^3$&-&$5.1\times 10^4$&-&&\\\hline\hline
  \end{tabular}
\end{table}

\begin{figure*}
  \includegraphics[width=12cm]{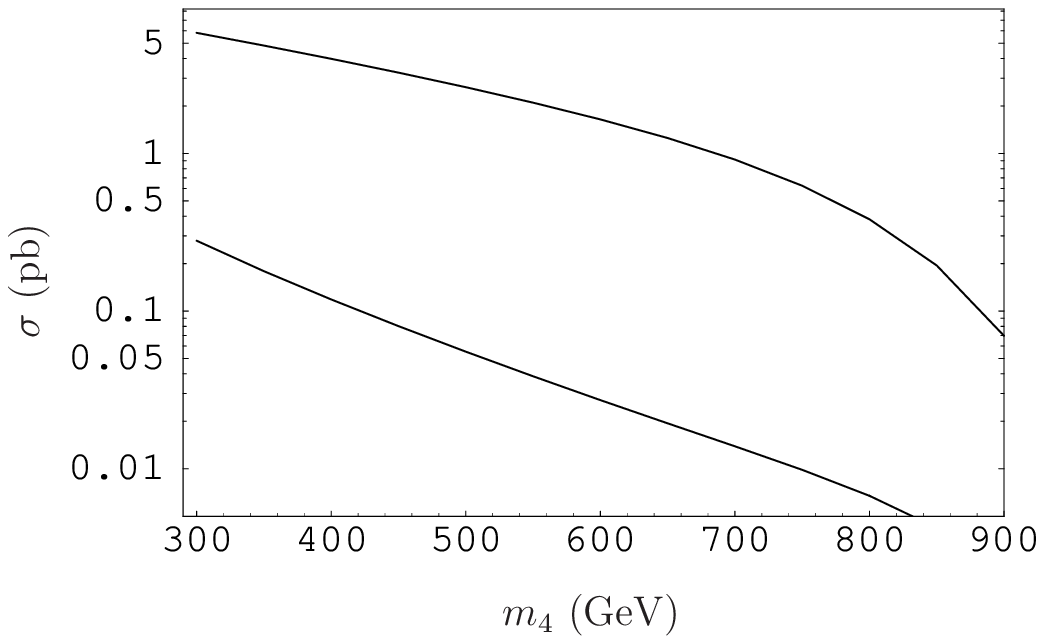}\\
FIG. 1: Photonic (lower line) and total (upper line) production
cross section for the FCNC single $u_4$ at$\sqrt{S}$=1 TeV as a
function of m$_4$ with $\kappa_{\gamma}=0.1$ and
$\kappa_{\gamma}=\kappa_{Z}=0.1$, respectively.
\end{figure*}

\begin{figure}

\includegraphics[width=12cm]{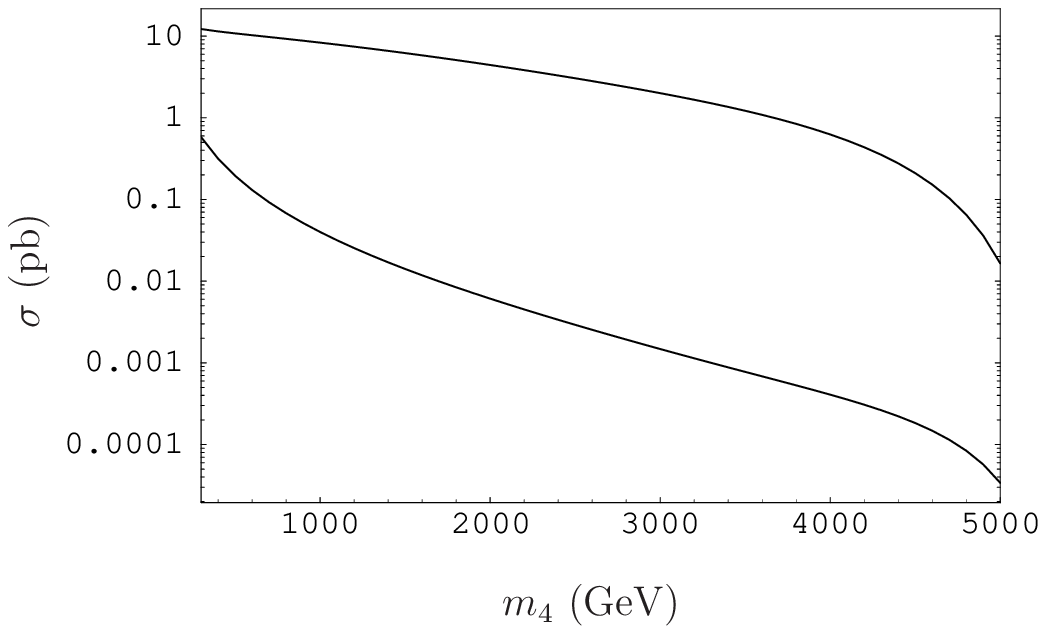}\\
FIG. 2: Photonic (lower line) and total (upper line) production
cross section for the FCNC single $u_4$ at $\sqrt{S}$=5.3 TeV as a
function of m$_4$ with $\kappa_{\gamma}=0.1$ and
$\kappa_{\gamma}=\kappa_{Z}=0.1$, respectively.
\end{figure}

\end{document}